\documentclass[prl,showpacs,twocolumn]{revtex4}

\usepackage{epsfig}
\usepackage{color}

\begin{document}
\title{Magnetic-field-induced binding of few-electron systems in shallow quantum dots}
\author{B. Szafran, S. Bednarek}
\affiliation{Faculty of Physics and Applied Computer Science, AGH
University of Science and Technology, PL-30059 Krak\'ow, Poland}
\author{F.M. Peeters}
\affiliation{Departement Fysica, Universiteit Antwerpen,
Groenenborgerlaan 171, B-2020 Antwerpen, Belgium}

\date{\today}

\begin{abstract}
Binding of few-electron systems in two-dimensional potential
cavities in the presence of an external magnetic field is studied
with the exact diagonalization approach. We demonstrate that for
shallow cavities the few-electron system becomes bound only under
the application of a strong magnetic field. The critical value of
the depth of the cavity allowing the formation of a bound state
decreases with magnetic field in a non-smooth fashion, due to the
increasing angular momentum of the first bound state. In the high
magnetic field limit the binding energies and the critical values
for the depth of the potential cavity allowing the formation of a
bound system tend to the classical values.

\end{abstract}


\maketitle Quantum dots \cite{jacak} (QDs) formed by weak
confinement potentials \cite{zhitenev} are susceptible to
perturbations by external charge defects and interface roughness
\cite{rough}. The spectra of disordered QDs
\cite{zhitenev,canalipairing} exhibit complex behavior in an
external magnetic field including pairing and bunching of the
charging lines, indicating the formation of multiple charged puddles
in local potential minima inside the large QD area. In the quantum
Hall effect disorder is held responsible for the formation of
localized electron reservoirs that allows for the pinning of the
Fermi level between the Landau levels \cite{data}. Coulomb blockade
of charge confined within these tiny reservoirs in quantum Hall
systems have recently been demonstrated \cite{ilani} through
scanning probe measurements. In this paper we consider electrons
localized in a nanometer-size reservoir created by a potential
fluctuation. The purpose of the present work is to determine the
role of the magnetic field in the formation of bound few-electron
states.

The present study is related to the problem of the existence of
bound excited states of the negative hydrogen ion $H^-$  or charged
donor center $D^-$ in a semiconductor. In the absence of a magnetic
field the $H^-$ has only one bound state \cite{Hill}, but an
infinite number of weakly bound and arbitrarily extended excited
states appears in weak magnetic field ($B\rightarrow 0 ^+$) due to
the interplay of the magnetic field induced localization and the
attractive force exerted on the extra electron by the neutral atom
$H$, or donor $D^0$ \cite{Avron}. This force results from the
electrostatic potential of $D^0$ which is attractive everywhere
\cite{Avron}. In two-dimensional quantum wells the $D^0$ potential
for the extra electron is repulsive at large distances \cite{Larsen}
which prevents the formation of extended excited bound states in the
weak magnetic field limit. For off-center $D^-$ in quantum wells
magnetic field induces angular momentum transitions are found
\cite{riva}, well-known from the theory of QDs \cite{manirev}.
States with angular momenta $L>4$ cease to be bound which results in
magnetic field induced $D^-$ evaporation \cite{riva}.

In the symmetric gauge the Hamiltonian of the electron in the
magnetic field $(0,0,B)$ contains a diamagnetic term
$V_d(x,y)={m\omega_c^2(x^2+y^2)}/{8}$ similar to the two-dimensional
harmonic oscillator potential with the cyclotron frequency
$\omega_c=eB/m$. Due to the parabolic form of $V_d$ one often
assumes \cite{Laughlin} that a strong magnetic field overcomes the
Coulomb repulsion between the electrons and makes them orbit around
the center of mass even in the absence of an external confinement
potential. Below we demonstrate, that although indeed the magnetic
field facilitates the binding of few-electron states, even in the
$B\rightarrow\infty$ limit a finite external potential is necessary
to stabilize the system. Moreover, we show that the potential
parameters stabilizing the system in the $B\rightarrow\infty$ limit
can be determined using classical physics.

For the purpose of the present study we need a confinement model
realistic enough to account for the unbound states. We choose the QD
model of a shallow two-dimensional Gaussian well. The Gaussian
potential is not only the simplest choice but is also realized in
vertical QDs \cite{Ashoori} with small \cite{Bednarek} gates. The
discussed physics is not limited to a Gaussian potential and is
independent of the specific choice of the confinement provided that
it has a finite size and depth. A weak magnetic field induces an
infinite number of single-electron states weakly bound in the
cavity. On the other hand the binding of extended two-electron
states at weak magnetic fields is excluded since the second electron
outside the cavity perceives only the Coulomb repulsion of the
electron bound in it. In this paper we look for the effect of the
magnetic field that overcomes the Coulomb blockade and allows
binding of additional electrons in the cavity. This effect can, for
example, be used to entangle spins \cite{divi} of the electrons
localized in the cavity and the unbound electrons.

Hamiltonian of a single electron with the magnetic field $(0,0,B)$
perpendicular to the plane of confinement reads (symmetric gauge), $
H=-\hbar^2\nabla^2/2m
+V_d(x,y)-\frac{1}{2}i\hbar\omega_c(x\partial/\partial
y-y\partial/\partial x)+V(x,y), $ where $m$ stands for the effective
band mass (GaAs material parameters are used) and $V(x,y)=-V_0
\exp(-(x^2+y^2)/R^2)$ is the potential well of depth $V_0$ and
radius $R$ (for the numerical calculations we take $R=50$ nm). In
the absence of the confinement ($V=0$) the diamagnetic term
$V_d(x,y)$ induces localization of the single-electron wave
functions, but by itself it can {\it not} keep several electrons
together, since the Hamiltonian eigenstates can be localized
anywhere. In particular the lowest Landau level (LLL) eigenfunction
(with eigenenergy $E_{LLL}=\hbar\omega_c/2$) can be put in the form
\cite{Maki}
$\Psi_{LLL}^{(X,Y)}=(\alpha/2\pi)^{1/2}\exp(-(\alpha/4)((x-X)^2+(y-Y)^2)+ieB/2\hbar(xY-yX))$,
where $(X,Y)$ is the arbitrary center of the Landau orbit and
$\alpha=eB/\hbar$.

\begin{figure}[htbp]
\hbox{\epsfysize=50mm
                \epsfbox[26 178 555 660] {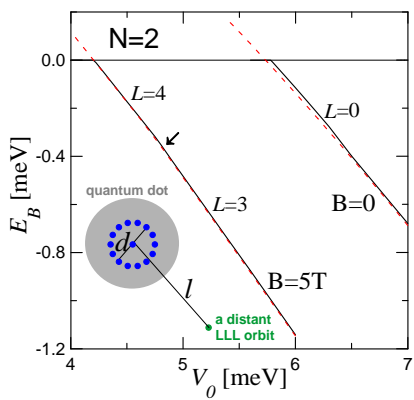}\hfill}

\caption{(color online) Binding energy of two-electron system vs
potential well depth for $B=0$ and 5T. Solid lines were calculated
using the multicenter basis (the inset shows the choice of centers,
$d$ is a variational parameter) and the dashed lines with the
finite-difference scheme. Arrow indicates where the ground state
angular momentum $L$ changes. }
\end{figure}

We studied the dissociation of the two-electron confined system
using the multicenter configuration interaction approach
\cite{aniso}, tailored in a way to account for both the bound and
unbound states of the two-electron system. The single-electron
states were obtained in the basis of the $\Psi_{LLL}^{(X,Y)}$ wave
functions \cite{aniso} in which the positions of the centers $(X,Y)$
as well as exponents $\alpha$ are treated as variational parameters.
The basis generates also the excited Landau levels and allows for a
precise solution of the few-electron Schr\"odinger equation
\cite{aniso}. We put 15 centers inside the dot (see: inset to Fig.
1) and a single center localized far away from the dot (at a
distance of $l=2$ cm). The method ejects an electron from the dot
into the LLL localized around the distant center when the
two-electron dot-confined system is not bound. For the unbound $N$
electron states the binding energy $E_B(N)=E_{N}-(E_{N-1}+E_{LLL})$
is not negative. Note, that the binding energy is the chemical
potential $\mu_N=E_N-E_{N-1}$ of the $N$-electron system determining
the single-electron charging of the QD devices calculated with
respect to the LLL $E_B(B)=\mu_N-E_{LLL}$.

The calculated binding energies are plotted in Fig. 1 by the thick
solid lines as function of the depth of the well. For $B=0$ the
two-electron system becomes bound in the $L=0$ state near $V_0=5.75$
meV. For 5T binding appears for a shallower well with $V_0=4.2$ meV,
and the first bound state has $L=4$.

To verify these results, we used an alternative approach \cite{dr}
using single-electron wave functions diagonalized numerically on a
finite-difference mesh. The method \cite{dr} allows for an exact
description of states localized in a circular potential of arbitrary
profile, but does not account for the delocalized LLL's since the
calculations are performed within a box of finite size. The results
are shown by the dashed lines in Fig. 1 and agree very well with our
previous approach. Above the dissociation threshold $E_B=0$ the
dashed lines become positive due to the absence of the distant LLL
in the basis \cite{dr}. The critical $V_0$ value is well reproduced.
For $B=0$ T, the finite difference approach works better than the
multicenter basis (see Fig. 1), because it better describes the
tails of the wave functions for weakly bound states, which in the
absence of a magnetic field can be arbitrarily long. In the
following we give the results as obtained with the finite difference
method \cite{dr}.

\begin{figure}[htbp]
\hbox{\epsfysize=60mm
                \epsfbox[12 181 555 687] {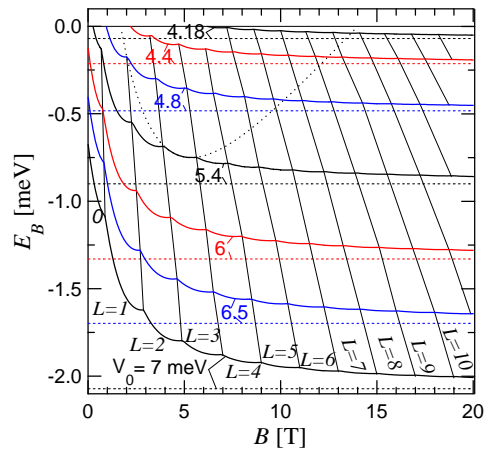}\hfill}

\caption{(color online) Binding energy of the two-electron system vs
the $B$ for depth of the potential from $V_0=7$ meV (lowest curve)
to 4.18 meV (highest curve). Horizontal dashed lines show the
binding energies of the classical system.  The energy cusps due to
the angular momentum transition are linked by the thin solid lines.
The dotted curve shows the binding energy of the state with $L=3$
for $V_0=5.4$ meV.}
\end{figure}

The two-electron binding energy as function of the magnetic field is
plotted in Fig. 2 for various depths of the dot.  The two electrons
are bound for any $B$ when $V_0=6$ meV, or more. The bound ground
state undergoes interaction-induced angular momentum transitions
\cite{manirev}. The interaction energy increases with increasing
magnetic field till it reaches a value at which the transition to a
more weakly localized state of lower interaction energy is more
favorable. At high magnetic fields the angular momentum transitions
\cite{manirev} keep the size of the system close to the size of its
classical counterpart \cite{bedanov}. The angular momentum
transitions are associated with the ground-state spin oscillations.
States with even $L$ correspond to spin singlets ($S=0$), and those
with odd $L$ to spin triplets ($S=1$).

For $V_0=5.4$ meV bound states appear only in the presence of an
external magnetic field (see Fig. 2). Notice that, for shallower QDs
the first bound state has non-zero $L$. We found that states of
specific $L$ are bound only within a finite interval of the magnetic
field (see the binding energy for $L=3$ at $V_0=5.4$ meV plotted by
the dashed line).

\begin{figure}[htbp] \hbox{\epsfysize=50mm
                \epsfbox[26 188 565 700] {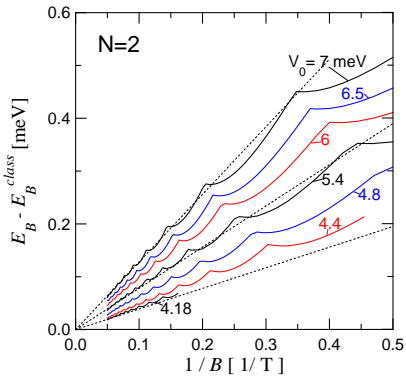}\hfill}

\caption{(color online) Binding energy of the two-electron system
minus the classical binding energy vs the inverse of the magnetic
field for depth of the potential from $V_0=7$ meV (highest curve) to
4.18 meV (lowest curve). The thin dashed straight lines are given by
 $0.39/B$ , $0.78/B$ and $1.28/B$.}
\end{figure}

\begin{figure}[htbp] \hbox{\epsfysize=60mm
                \epsfbox[22 3 550 530] {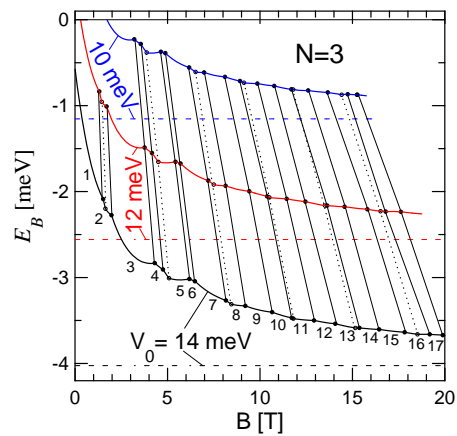}\hfill}

\caption{(color online) Binding energy of the three-electron system
vs the magnetic field for three values of the potential depth.
Horizontal dashed lines show the binding energies of the classical
system.  The energy cusps corresponding to the angular momentum
transitions in the three- (two-) electron ground state are linked by
 thin solid (dotted) lines. The numbers below the lowest curve
list $L$ for $N=3$.}
\end{figure}

\begin{figure}[htbp] \hbox{\epsfysize=50mm
                \epsfbox[16 126 573 650] {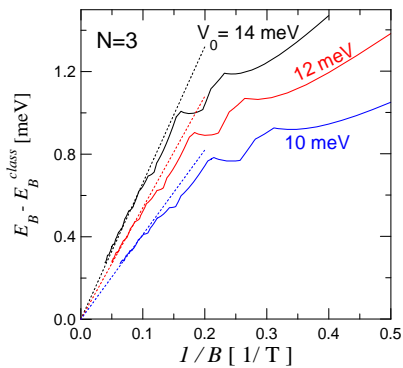}\hfill}

\caption{(color online) Same as Fig. 3 but for $N=3$.  The thin
dashed straight lines are given by  $6.6/B$, $5.4/B$ and $4.1/B$.}
\end{figure}

\begin{figure}[htbp] \hbox{\epsfysize=50mm
                \epsfbox[16 126 437 514] {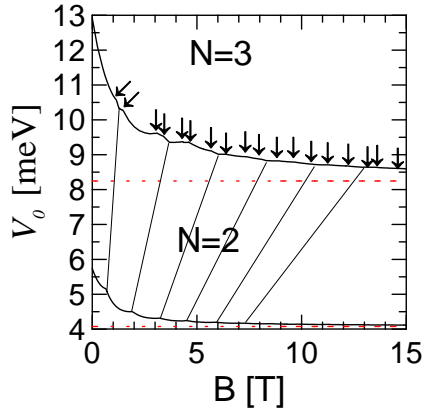}\hfill}

\caption{(color online) The maximum number of bound electrons on the
$V_0/B$ plane. Below the lowest curve only one electron can be
bound, and above the highest the third electron becomes bound,
between the curves the maximum number of bound electrons is two. The
horizontal dashed lines show the classical depths of the well
allowing binding of the second and third electron.}
\end{figure}

\begin{figure}[htbp] \hbox{\epsfysize=45mm
                \epsfbox[28 163 556 701] {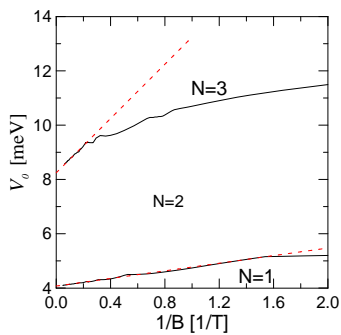}\hfill}

\caption{(color online) Same as Fig. 6 but in $1/B$ scale.  The thin
dashed straight lines are given by  $V_N^{class}+A_N/B$, with
$V_2^{class}=4.073$ meV, $V_3^{class}=8.245$ meV, $A_2=0.7$,
$A_3=5$.}
\end{figure}

The horizonal dashed lines in Fig. 2 show the values of the binding
energies obtained for classical point charges \cite{bedanov,Farias},
via minimization of the two-electron potential energy
$E_B^{class}=\frac{e^2}{4\pi\epsilon\epsilon_0}\frac{1}{w}-2V(w/2,0)$
with respect to the interelectron distance $w$. The quantum binding
energies tend to the classical values at high $B$. This is shown
clearly when we plot the difference between the quantum and
classical values as function of the inverse of the magnetic field
(see Fig. 3). We conclude that the binding energy tend as $1/B$ to
the classical value in the limit of high magnetic fields
\cite{EPJD}. The classical character of the binding is due to the
point localization of the single-electron LLL wave functions in the
high magnetic field.

The cusps of the three-electron binding energy
 are due to both the $N=3$ and
$N=2$ angular momentum transitions (cf. Fig. 4). The $1/B$
convergence of the binding energy to the classical limit is also
found for three electrons (see Fig. 5).

Fig. 6 shows the 'phase diagram' for the maximum number of bound
electrons as function of the well depth and the magnetic field. The
thin lines joining the cusps of the phase diagram boundaries
corresponding to $N=2$ angular momentum transitions. The cusps
related to the $N=3$ angular-momentum transitions are marked with
arrows. Dashed lines show the depth of the potential well which
allows binding of the second and third classical electron
\cite{Farias}. In Fig. 7 we plot the phase diagram as function of
$1/B$ to demonstrate the convergence of the phase boundaries to the
classical result at infinite magnetic field.

In an external magnetic field the diamagnetic effect increases
faster the free-electron LLL than the energies of the states
localized in the cavity. This leads to the energy shift of the
ionization threshold observed in a magnetic field. According to our
results the magnetic field induces the formation of bound electron
states in QDs. Such a formation was actually observed in
single-electron charging experiments on self-assembled QDs
\cite{Miller}. The QDs \cite{Miller} at low magnetic fields bind six
electrons. Only for $B>8$ T the capacitance signal related to the
binding of sixth and seventh electron emerge from the background of
the LLL wetting layer charging (see Fig. 3 of Ref.\cite{Miller}).
Self-assembled QDs have a fixed depth of the confinement potential,
but the phase diagram (Figs. 6 and 7) can be experimentally
determined in the capacitance measurement of the vertical quantum
dot device \cite{Ashoori}. The shape of the confinement potential is
exactly Gaussian for a proper choice of the size of the gate (see
Fig. 6 of \cite{Bednarek}) forming a small electrostatic QD. The
depth of the potential is simply changed by the gate voltage. The
number of bound electrons can be determined for any magnetic field
by counting the capacitance peaks appearing as function of the gate
voltage below the two-dimensional electron gas charging (like the
one of Fig. 3 Ref. \cite{Miller}).

The discussed effect can be used to entangle electron spins
\cite{divi}. One needs a cavity which at $B=0$ binds only a single
electron. The second electron is bound by the field and one can
arrange for the bound state to have zero spin (even $L$). After the
magnetic field is switched off the second electron will become
unbound and is repelled by the dot and thus goes with its spin state
entangled with the localized electron.

In summary, we have shown that a strong magnetic field induces the
formation of bound few-electron states in potential cavities that
are deep enough to bind the classical point electrons.

{\bf Acknowledgment} This paper was supported by the Foundation for
Polish Science, the Flemish Science Foundation and the
EU-NoE:SANDiE.

\end{document}